\documentclass[twocolumn,secnumarabic,amssymb, nobibnotes, aps, prb, superscriptaddress]{revtex4-2}
\usepackage{lipsum}
\usepackage{titlesec}
\titlespacing\section{0pt}{12pt plus 4pt minus 4pt}{1pt plus 20pt minus 2pt}
\usepackage{xcolor}
\usepackage{amsmath}
\usepackage{comment}
\usepackage{physics}
\usepackage{afterpage}
\usepackage{placeins}
\usepackage{graphicx}
\usepackage{float}
\usepackage{booktabs}
\usepackage{multirow}
\usepackage{array}
\usepackage{setspace}
\graphicspath{{Figs/}}
\usepackage{siunitx}
\usepackage{hhline}
\usepackage{xfrac}
\usepackage{float,graphicx}
\usepackage{mathtools}
\usepackage{listings}
\usepackage{amssymb}
\usepackage{titlesec}
\usepackage{amsfonts}
\usepackage[version=4]{mhchem}

\usepackage{epstopdf}
\catcode`@11
\def\seceqaa{\@addtoreset{equation}{section}
\def\theequation{A\arabic{equation}}}
\def\seceqbb{\@addtoreset{equation}{section}
\def\theequation{B\arabic{equation}}}
\def\seceqcc{\@addtoreset{equation}{section}
\def\theequation{C\arabic{equation}}}
\def\seceqdd{\@addtoreset{equation}{section}
\def\theequation{D\arabic{equation}}}
\def\seceqee{\@addtoreset{equation}{section}
\def\theequation{E\arabic{equation}}}
\def\seceqff{\@addtoreset{equation}{section}
\def\theequation{F\arabic{equation}}}
\def\seceqgg{\@addtoreset{equation}{section}
\def\theequation{G\arabic{equation}}}
\def\seceqhh{\@addtoreset{equation}{section}
\def\theequation{H\arabic{equation}}}
\catcode`@11

\begin{document}

\title{ Anomalous Hall effect from gapped nodal line in Co$_2$FeGe Heusler compound} 

\author{Gaurav K. Shukla}
\affiliation{School of Materials Science and Technology, Indian Institute of Technology (Banaras Hindu University), Varanasi 221005, India}

\author{Jyotirmay Sau}
\affiliation{S. N. Bose National Centre for Basic Sciences, Kolkata 700098, West Bengal, India}

\author{Nisha Shahi}
\affiliation{School of Materials Science and Technology, Indian Institute of Technology (Banaras Hindu University), Varanasi 221005, India}

\author{Anupam K. Singh}
\affiliation{School of Materials Science and Technology, Indian Institute of Technology (Banaras Hindu University), Varanasi 221005, India}

\author{Manoranjan Kumar}
\affiliation{S. N. Bose National Centre for Basic Sciences, Kolkata 700098, West Bengal, India}

\author{Sanjay Singh}
\affiliation{School of Materials Science and Technology, Indian Institute of Technology (Banaras Hindu University), Varanasi 221005, India}
%\date{today}

\begin{abstract}
%In recent past, Weyl semimetals (WSMs) got ample attention due to their non-trivial topological band structure. Many 
Full Heusler compounds with Cobalt as a primary element show anomalous transport properties owing to the  Weyl fermions and broken time-reversal symmetry. We present here the study of anomalous Hall effect (AHE) in Co$_2$FeGe  Heusler compound. The experiment reveals anomalous Hall conductivity (AHC) $\sim 100\,S/cm$  at room temperature with an intrinsic contribution of $\sim 78\,S/cm$ . The analysis of anomalous Hall resistivity suggests the scattering independent intrinsic mechanism dominates the overall behaviour of anomalous Hall resistivity. The first principles calculation reveals that the Berry curvature originated by gapped nodal line near E$_F$ is the main source of AHE in Co$_2$FeGe Heusler compound. The  theoretically calculated AHC is in agreement with the experiment.  
\end{abstract}

\maketitle
%\section{Introduction}
%\vspace*{-3mm}

Weyl semimetals (WSMs) host exotic transport properties resulting from their non-trivial topological band structure \cite{wang2017quantum,nagaosa2020transport,yan2017topological,hsieh2009observation,xu2015ultrasensitive,hosur2013recent}. WSMs are characterised by chiral anomaly and linear band crossing points known as Weyl points or Weyl nodes \cite{lv2021experimental}.  The existence of Weyl nodes are possible in metal or semimetal with  broken inversion symmetry (IS) and/or time reversal symmetry (TRS). These broken symmetries lift the two fold degeneracy of electronic  bands %enforced by 
in the framework of Kramer's theorem\cite{klein1952degeneracy} and the linearly dispersing touching points of two non-degenerate  bands become Weyl points \cite{chen2021anomalous,burkov2016topological}. The Hamiltonian of the system describing Weyl nodes can be written in term of basis vector of three Pauli matrices and hence any perturbation with linear combination of Pauli matrices can not destroy the Weyl nodes \cite{chen2021anomalous}.  Also no  other symmetries require (except translational symmetry) for the protection of  Weyl nodes that represent the Weyl nodes as topologically stable object \cite{yang2016dirac}. 
%~In 1984 Michel Berry proposed that energy level crossings of a Hamiltonian act as magnetic mono-pole and introduced the concept of Berry curvature\cite{berry1984quantal}, which is equivalent to pseudo-magnetic field in momentum space  results in various intriguing phenomenon in material such as anomalous Hall effect (AHE), anomalous Nernst effect (ANE), chiral magnetoresistance, and second harmonic generation etc\cite{nagaosa2020transport,noky2018characterization,noky2019large}. 
Michel Berry introduced the concept of Berry curvature \cite{berry1984quantal} which may be mapped to  pseudo magnetic field and the degenerate Weyl points %can form source and sink of  Berry curvature or magnetic monopole. 
~correspond to quantized monopoles form source and sink of Berry curvature.
~These concepts help us to understand the various intriguing phenomenon like anomalous Hall effect (AHE) \cite{manna2018heusler,noky2018characterization}, anomalous Nernst effect (ANE) \cite{manna2018colossal}, chiral magnetoresistance \cite{nagaosa2020transport,lv2021experimental} and second harmonic generation \cite{nagaosa2020transport} etc. %Since, Weyl points belong to such crossings hence act as source and drain of the Berry curvature.  
%IS breaking 
Weyl points induced by the breaking of IS was observed experimentally first time in TaAs \cite {xu2015discovery,yang2015weyl,lv2015observation} and investigated extensively in its family members \cite{PhysRevLett.117.146403,xu2015experimental}  and also in other materials   \cite{soluyanov2015type,sun2015prediction,liu2014weyl}.  TRS breaking WSMs are also known as magnetic WSMs discovered very recently and created much interest due to  exhibiting large intrinsic AHE and anomalous Hall conductivity  (AHC) which is proportional to the separation of Weyl nodes \cite{li2020giant, manna2018heusler,zyuzin2012weyl}. The magnetic WSMs have added advantage over conventional WSMs because an external magnetic field can be used to manipulate the properties of magnetic WSMs \cite{destraz2020magnetism}.
 Co$_3$SnS$_2$, a ferromagnet kagome lattice is firstly discovered as  magnetic WSM exhibits large Berry curvature resulting the giant intrinsic AHC 1130 S/cm, which is an order of magnitude larger than the typical ferromagnets\cite{liu2018giant}.  This large Berry curvature attribute to the presence of Weyl nodes and nodal rings of linear crossings in the spin-up channel based on band inversion. Besides, Co$_3$SnS$_2$, several other materials such as  pyrochlore iridates \cite{wan2011topological,ueda2018spontaneous,takahashi2018anomalous}, PrAlGe  \cite{destraz2020magnetism}, YMnBi$_2$ \cite{borisenko2019time}, Mn$_3$Sn \cite{kuroda2017evidence},  Co$_3$MM'X$_2$ ( M/M' = Ge, Sn, Pb, X=S, Se, Te )  \cite{luo2021cobalt} and Heusler alloys  \cite{belopolski2019discovery,li2020giant,chang2016room,hirschberger2016chiral,shi2018prediction}  have been identified as magnetic WSMs theoretically and/or  experimentally. 
 
Among various magnetic WSM candidates, % materials Heusler based magnetic WSMs are promising materials due to their extensive tunability\cite{li2020giant,wollmann2017heusler}.  
Co$_2$ based Heusler compounds have been found to be more interesting that drives the large AHC due to a large Berry curvature \cite{belopolski2019discovery,chang2016room,li2020giant,dulal2019weak}. In most of the Co$_2$ based magnetic WSMs, gapped nodal line  in momentum space has been found as a major source of Berry curvature and creates intrinsic AHE \cite{ernst2019anomalous,guin2019anomalous,li2020giant}. 
In  these compounds, three gapless nodal lines which are protected by three mirror planes exist  in the absence of magnetization. %\textcolor{red}{In 3-dimension the curve where the two bands cross each other along a closed curve is called nodal line\cite{fang2016topological}}. 
The gapless nodal line gaps out with introduction of spin orbit coupling(SOC) according to the magnetization direction \cite{manna2018heusler,ernst2019anomalous}. %The gapped nodal lines  act as a source of Berry curvature and create large intrinsic AHE \cite{noky2018characterization,ernst2019anomalous}. 
For example, Co$_2$MnGa exhibits a giant AHC around 1260 S/cm at 60\,K due to large Berry curvature associated with the gapped nodal line \cite{guin2019anomalous}. Co$_2$VGa  exhibit the AHC around  140 S/cm due to slight gapped nodal line result from reduced mirror symmetry upon introducing magnetization  \cite{manna2018colossal}. Recently, Co$_2$MnAl Weyl semimetal reported to have large AHC 1600 S/cm at 2\,K due to gapped nodal ring in momentum space which  is as large as for 3D quantum AHE \cite{li2020giant}.

Theoretical investigation performed on the Co$_2$FeGe Heusler compound reveals the existence of nodal line  above the Fermi energy (E$_F$) results in to an intrinsic AHE \cite{noky2018characterization,noky2020giant}. 
In this manuscript, we present study of AHE on Co$_2$FeGe Heusler compound using both experiment and theoretical calculations.  We found the experimental AHC $\sim$ 100 S/cm at 300\,K and shows weak dependence on temperature.
%nearly insensitive to the temperature from {several K} to 300 K. 
Our first principles calculation reveals that the magnetization induced gapped nodal line near the E$_F$ is the main source of AHE in Co$_2$FeGe.

%%%%%%%%%%%%%%%%%%%%%%%%%%%%%%%%%%%%%%%%%%%
\begin{figure}[htp]
\includegraphics[width=0.5\textwidth]{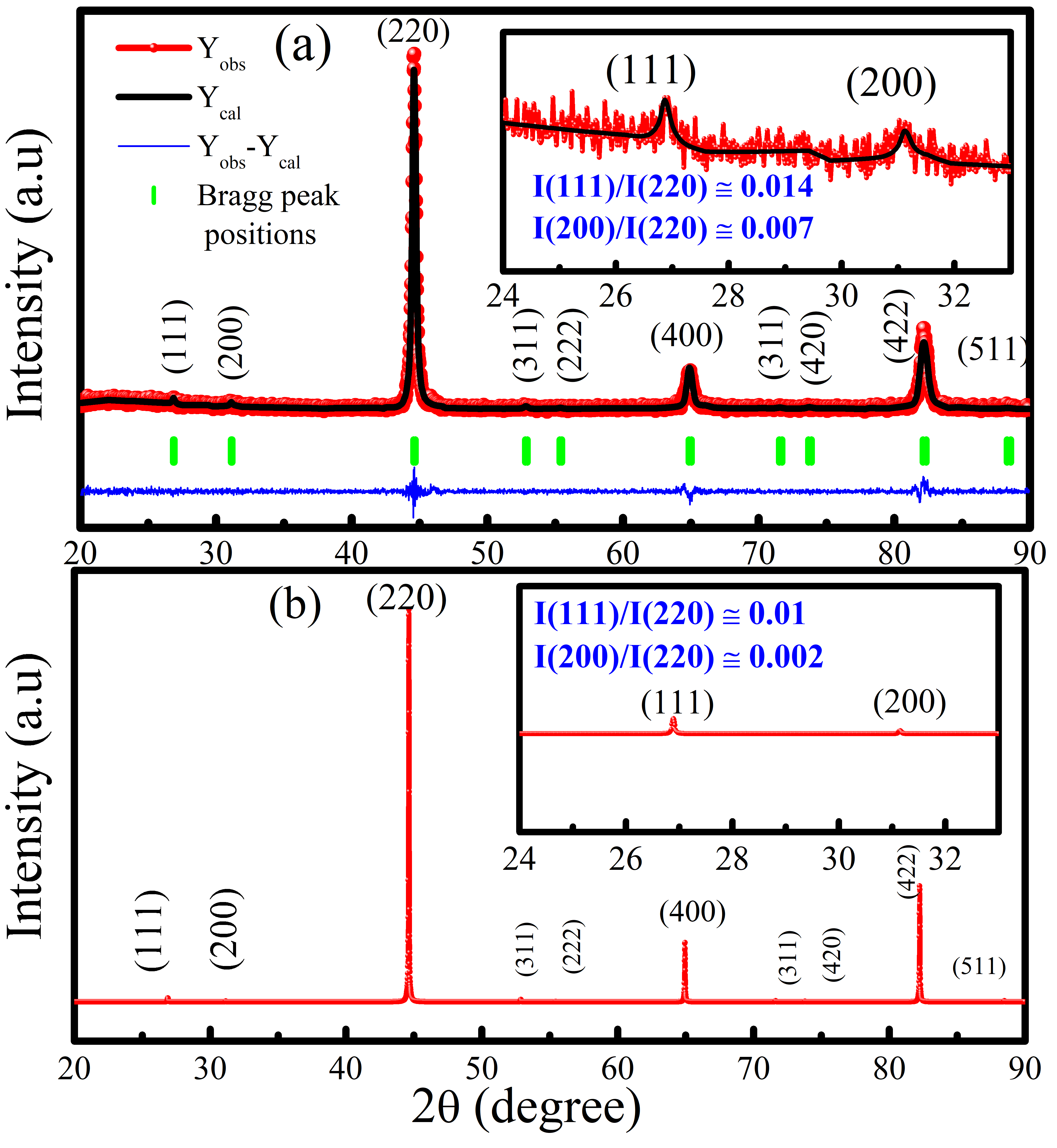}
\caption[Content]{(a)Rietveld modeling of X-ray diffraction pattern of Co$_2$FeGe  at room temperature.(b) Simulated XRD pattern of Co$_2$FeGe. Inset of figures show the enlarged view around the (111) and (200) superlattice reflections.}
  \label{Fig1}
\end{figure}
%%%%%%%%%%%%%%%%%%%%%%%%%%%%%%%%%%%%%%%%%%%

%\section{Experiment and method}
%\vspace*{-3mm}
Polycrystalline Co$_2$FeGe compound was synthesized by standard arc melting technique \cite{tsai2015vacuum} in the presence of pure argon atmosphere using 99.99\% pure individual elements. The sample was remelted several times for the homogeneous mixing of involved elements. A small weight loss of 0.62 \% was notified after melting. Small piece was taken from sample and crushed into powder for X-ray diffraction measurement. The polished rectangular piece of the dimension $4\times2\times0.65$ mm$^3$  was used for  temperature and magnetic field dependent transport measurements %to obtain the longitudinal resistivity and the Hall resistivity respectively 
using cryogen free measurement system (Cryogenic, CFMS). To obtain the actual transverse resistivity($\rho\textsubscript{H}$), raw Hall resistivity data ($\rho\textsubscript{H}^{raw}$) was anti-symmetrized  by averaging  the difference of ${\rho\textsubscript{H}^{raw}}$ at the positive field and negative field with respect to the field sweep direction.

The electronic band structure and magnetic properties of the Co$_2$FeGe are calculated employing density functional theory (DFT) using the Vienna-ab initio simulation package (VASP) \cite{hafner2008ab}. Exchange-correlation potential is approximated with generalized gradient approximation and projector augmented wave method (PAW)\cite{blochl1994projector} is used for core-valence interaction. The calculations are performed with K-mesh of 10×10×10 for the Fm$\bar{3}$m space group(space group no.225). The plane-wave basis is used with cut-off energy 500 eV and force convergence for the optimisation is kept below 0.001 eV/\si{\angstrom}. Self-consistent calculations are performed to get the charge density and thereafter the band structure is calculated. To  understand the DFT band structure, Wannier interpolated bands and corresponding tight-binding parameters are calculated using Wannier90~\cite{Pizzi_2020,marzari_97}. The anomalous Hall conductivity (AHC), Berry curvature and energy gap for the given parameters are calculated using the  WannierTool~\cite{WU2018405}. For AHC calculation Kubo formalism is used in the clean limit~\cite{Gradhand_2012}. Co$_2$FeGe material possess Fm$\bar{3}$m symmetry and have three relevant mirror planes \textit{m}$_x$(\textit{k}$_x$=0), \textit{m}$_y$(\textit{k}$_y$=0), \textit{m}$_z$(\textit{k}$_z$=0) and three $C_4$ rotation axes \textit{k}$_x$,\textit{k}$_y$ and \textit{k}$_z$ \cite{chang}. The magnetization is oriented along the z-axis and the SOC is also considered along the same axis. 

%%%%%%%%%%%%%%%%%%%%%%%%%%%%%%%%%%%%%%%%%%%
\begin{figure}[t]
    \centering
    \includegraphics[width=0.5\textwidth]{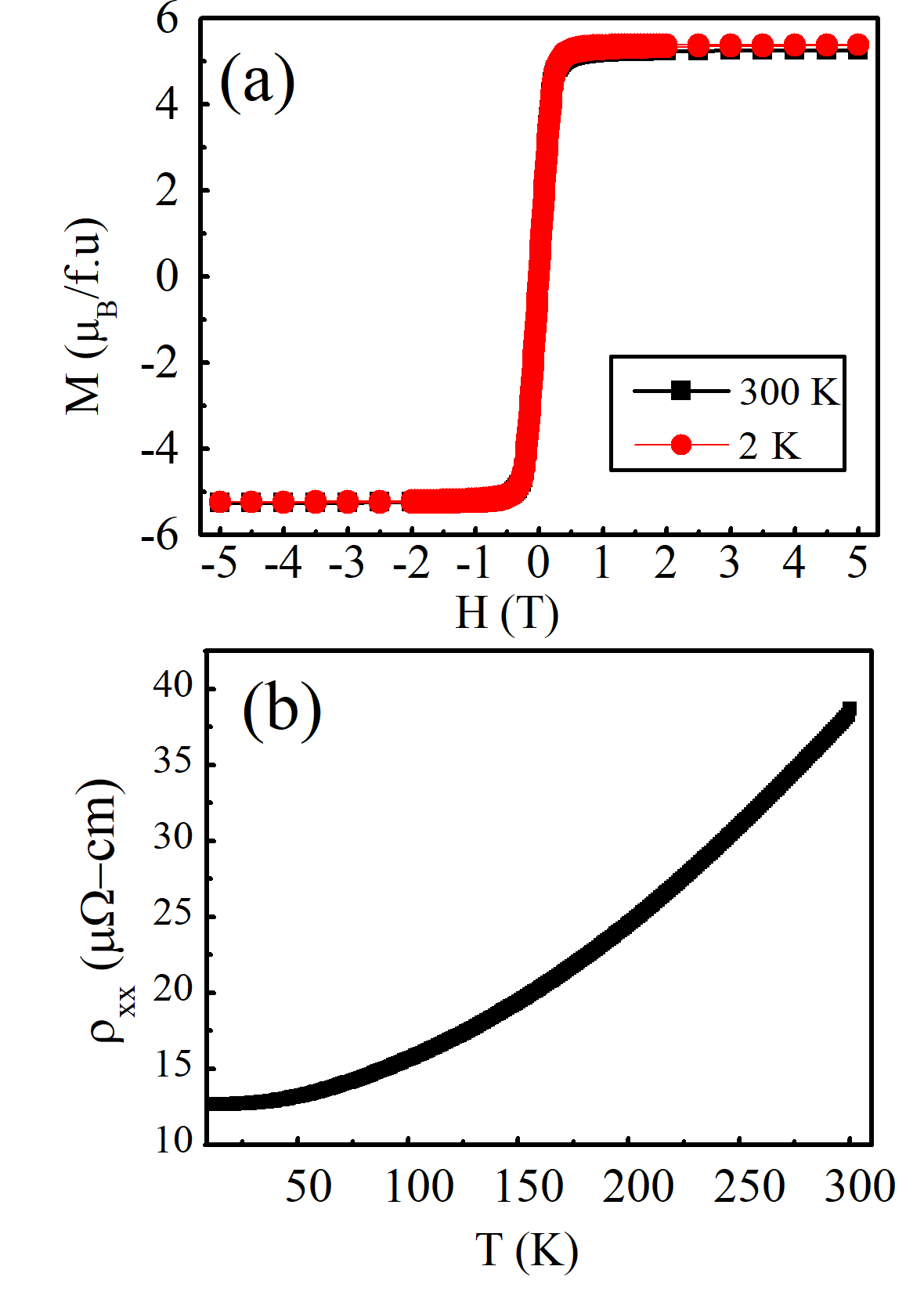}
    \caption{(a) The field dependent magnetization curves at 2\,K and 300\,K. (b)Temperature dependent longitudinal resistivity $\rho\textsubscript{xx}$. }%Inset fig.(i), (ii), and (iii) show resistivity data from 10 K to 50 K, 50K to 250K and 250 K  to 300K respectively (black curves). Fitting is shown by red curve.
 
    \label{Fig2}
\end{figure}

%%%%%%%%%%%%%%%%%%%%%%%%%%%%%%%%%%%%%%%%%%%

%%%%%%%%%%%%%%%%%%%%%%%%%%%%%%%%%%%%%%%%%%%
\begin{figure*}[t]
    \centering
    \includegraphics[width=1\textwidth]{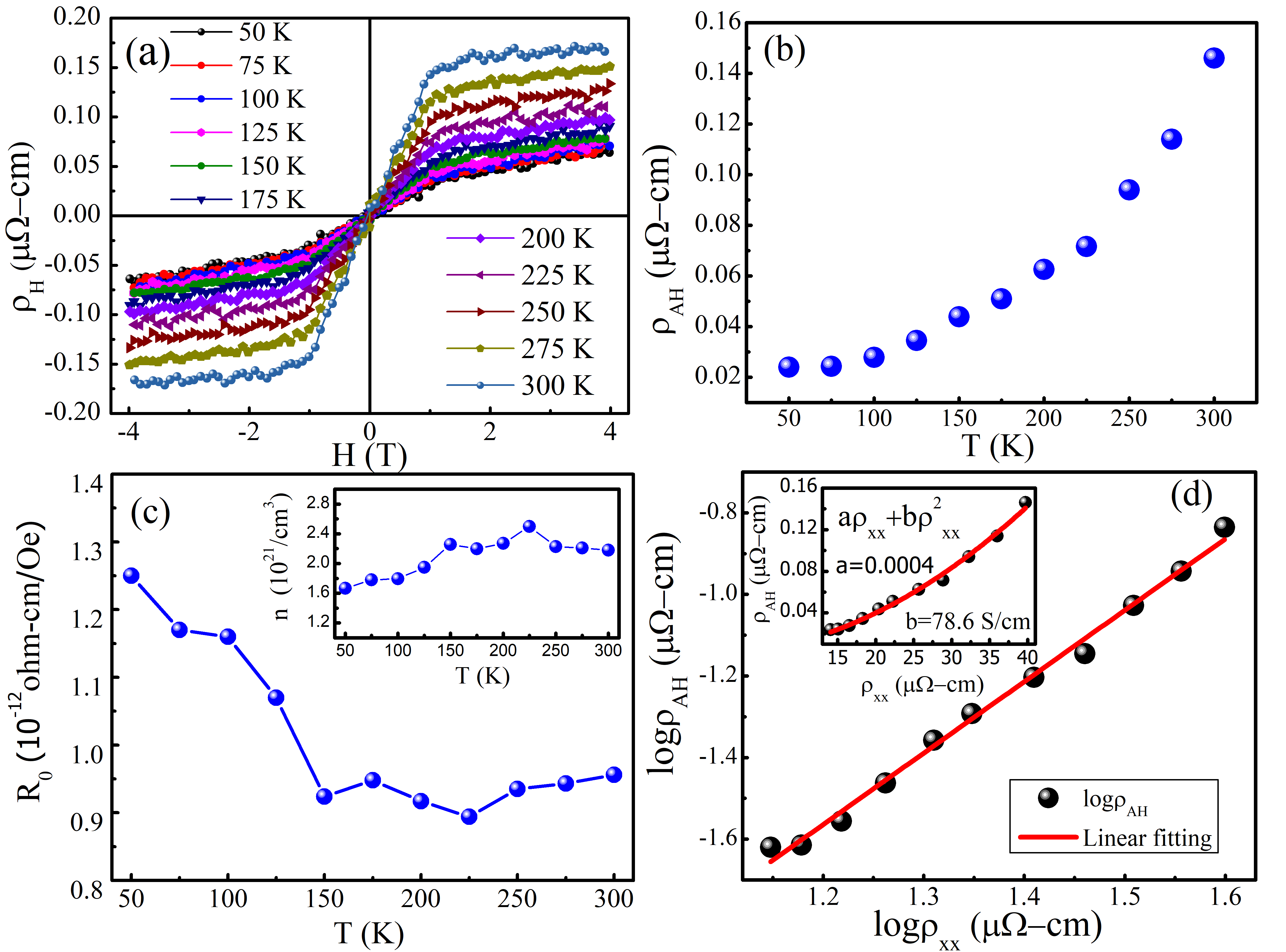}
    \caption{(a) Field dependent Hall resistivity$\rho\textsubscript{H}$ at indicated temperatures. (b) Temperature dependent anomalous Hall resistivity $\rho\textsubscript{AH}$. (c) Temperature dependent normal Hall coefficient R$_0$. Inset shows the temperature dependent carrier concentration n. (d) Double logarithmic plot between $\si{\rho}\textsubscript{AH}$ and \si{\rho}\textsubscript{xx} (black balls) and linear fitting is shown by red line. Inset  shows the graph between \si{\rho}\textsubscript{AH} and \si{\rho}\textsubscript{xx} (black balls) and fitting using Eq.(1) is shown by red line.}
    \label{fig3}
\end{figure*}

%\section{Result and discussion}
%\vspace*{-3mm}
%\subsection{Structural properties}
%\vspace*{-3mm}
X-ray diffraction (XRD) pattern of the sample collected at room temperature for structural investigation and phase purity. The Rietveld analysis of the XRD pattern was done using Fullprof software \cite{rodriguez2001introduction}. The space group Fm$\bar{3}$m and Wyckoff positions: 8c (1/4, 1/4, 1/4) occupied by Co atoms, whereas 4b (1/2, 1/2, 1/2) and 4a (0, 0, 0) occupied by Fe and Ge atoms, respectively were used. The observed XRD patterns depicted in Fig.\ref{Fig1}a show that all the Bragg peaks observed are well indexed confirming the phase purity (cubic) of the Co$_2$FeGe sample. 
The refined unit cell parameter was found 5.74 \AA.

For full Heusler alloys the reflections index relation h, k and l = odd number or (h+k+l)/2 = (2n+1) are the superlattice reflection, while (h+k+l)/2 = 2n are the fundamental reflections\cite{takamura2009analysis,sakuraba2020giant}. 
The ordered structure of full Heusler alloys (L2$_1$ structure) generally mark the presence of (111) and (200) superlattice reflections; presence of (111) peak indicates the chemical ordering of atom at the octahedral position, and (200) peak indicates the ordering at the tetrahedral position, while the intensity of (220) fundamental reflection is independent from atomic ordering \cite{wurmehl2006investigation}. 
The presence of both (111) and (200) superlattice peaks primarily suggest the L2$_1$ structure of Co$_2$FeGe as shown in inset of Fig.\ref{Fig1}a. To compare the experimentally observed relative intensities of the superlattice reflections with theory, we simulated the  XRD pattern using PowderCell software as shown in Fig.\ref{Fig1}b. Similar to the observed, weak intensities of (111) and (200) superlattice reflections have also been observed in simulated XRD pattern. The enlarged view of simulated superlattice peaks are shown in the inset of Fig.\ref{Fig1}b. The weak intensity of superlattice reflections is due to the small difference between atomic scattering factor of constituent 3d metals of Co$_2$FeGe \cite{balke2007structural,balke2008rational}. The measured and simulated $\frac{I_{111}}{I_{220}}$ and $\frac{I_{200}}{I_{220}}$ are given in inset of Fig.\ref{Fig1}a and Fig.\ref{Fig1}b respectively. A good match between the measured and simulated XRD pattern suggests the formation of ordered structure of Co$_2$FeGe. 

%\subsection{Magnetic measurement}
%\vspace*{-3mm}
Magnetic isotherms up to field of 5 Tesla were recorded at temperatures 2\,K and 300\,K depicted in Fig.\ref{Fig2}a.The magnetic moment was found to be  5.38 $\mu_B$/f.u and 5.24 $\mu_B$/f.u at 2\,K and 300\,K, respectively. The observed magnetic moment is well agreement with the literature \cite{rai2012comparative,amari2016theoretical} as well as our theoretical calculation (discussed later).
%theoretically reported value 5.39 $\mu_B$/f.u using LSDA \cite{rai2012comparative} and 5.32  $\mu_B$/f.u using GGA approximation\cite{amari2016theoretical}.
~The variation of longitudinal resistivity ($\rho\textsubscript{xx}$) as a function of temperature  from 10\,K to 300\,K is shown in Fig.\ref{Fig2}b. The $\rho\textsubscript{xx}$ increases with increasing temperature and a  residual resistivity about 14 $\mu\Omega$cm is observed. The non-linear behaviour of resistivity above 50\,K suggests the combined phonon and magnon scattering state \cite{goodings1963electrical}. %As temperature declines the phonon contribution subsides, so  electron-electron and/or electron-magnon scattering state can be attributed below 50 K.  
~The residual resistance ratio (RRR= $\rho\textsubscript{xx}(300 K)$/$\rho\textsubscript{xx} (10 K)$)
which quantifies the degree of disorder is 2.82. This value is larger than the most  of Co$_2$-based Heusler alloy \cite{markou2019thickness,imort2012anomalous,vidal2011influence,roy2020anomalous,ernst2019anomalous} signifies comparatively clean sample.
%\subsection{Hall measurement}
%\vspace*{-3mm}

After investigation of the phase purity, saturation magnetization and temperature variation of the longitudinal resistivity, we carried out a detailed magneto-transport measurement in a wide range of temperature 50 K to 300 K to study the AHE in the Co$_2$FeGe Heusler alloy. Hall resistivity (\si{\rho}\textsubscript{H}) curves measured up to 4 Tesla magnetic field at different temperatures are shown in Fig.\ref{fig3}a. The \si{\rho}\textsubscript{H}  is given by equation $\rho\textsubscript{H}(H)$  =  R$_0$H + R$_s$M \cite{nagaosa2006anomalous,ernst2019anomalous}.
Where R$_0$ and R$_s$ are the normal and anomalous Hall coefficients. H is the external applied magnetic field and M is magnetization of the material. 
It is evident from  Fig.\ref{fig3}a that \si{\rho}\textsubscript{H} initially increases at lower fields indicating AHE in Co$_2$FeGe and shows the positive slope according to sign of ordinary Hall coefficient at higher field region for all temperatures.
The anomalous Hall resistivity (\si{\rho}\textsubscript{AH}) calculated by the zero field extrapolation of high field Hall data with the ordinate, which is equivalent to Hall voltage response arising due to spontaneous magnetization in the absence of external magnetic field. Fig.\ref{fig3}b, shows the extracted \si{\rho}\textsubscript{AH} versus temperature plot, which displays that the Hall resistivity increases with increasing temperature and acquire a maximum value of 0.14$\mu\Omega$-cm at room temperature and a small value 0.02 $\mu\Omega$-cm at 50K.
The slope of high field Hall resistivity data gives  the normal Hall coefficient (R$_0$) and variation of  R$_0$ with temperature is shown in Fig \ref{fig3}c.  By using relation  R$_0$=$\frac{1}{ne}$, we calculated carrier density (n) and plotted in the inset of Fig.\ref{fig3} (c) with temperature. The value of n was found $\sim$ 2$\times$10$^{21}$/cm$^3$. The positive value of R$_0$ indicates that the  holes are majority charge carriers in the whole temperature range.

%%%%%%%%%%%%%%%%%%%%%%%%%%%%%%%%%%%
\begin{figure}[t]
    \centering
    \includegraphics[width=0.5\textwidth]{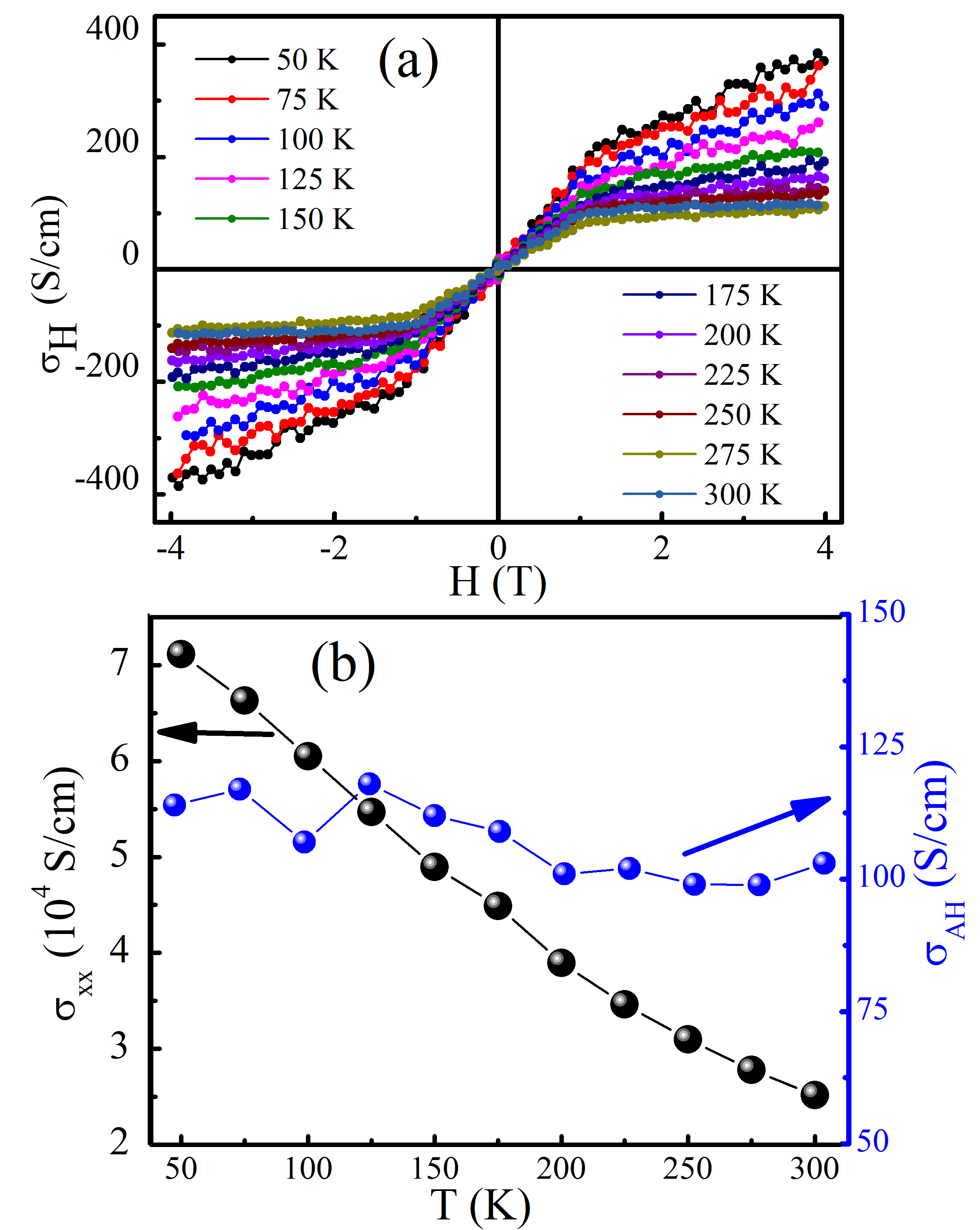}
    \caption{(a) Field dependent Hall conductivity $\sigma\textsubscript{H}$ (b) Temperature dependent longitudinal conductivity $\sigma\textsubscript{xx}$(black spheres) and AHC  (blue spheres).}
    \label{fig4}
\end{figure}
%%%%%%%%%%%%%%%%%%%%%%%%%%%%%
\begin{figure}
\includegraphics[width=\columnwidth]{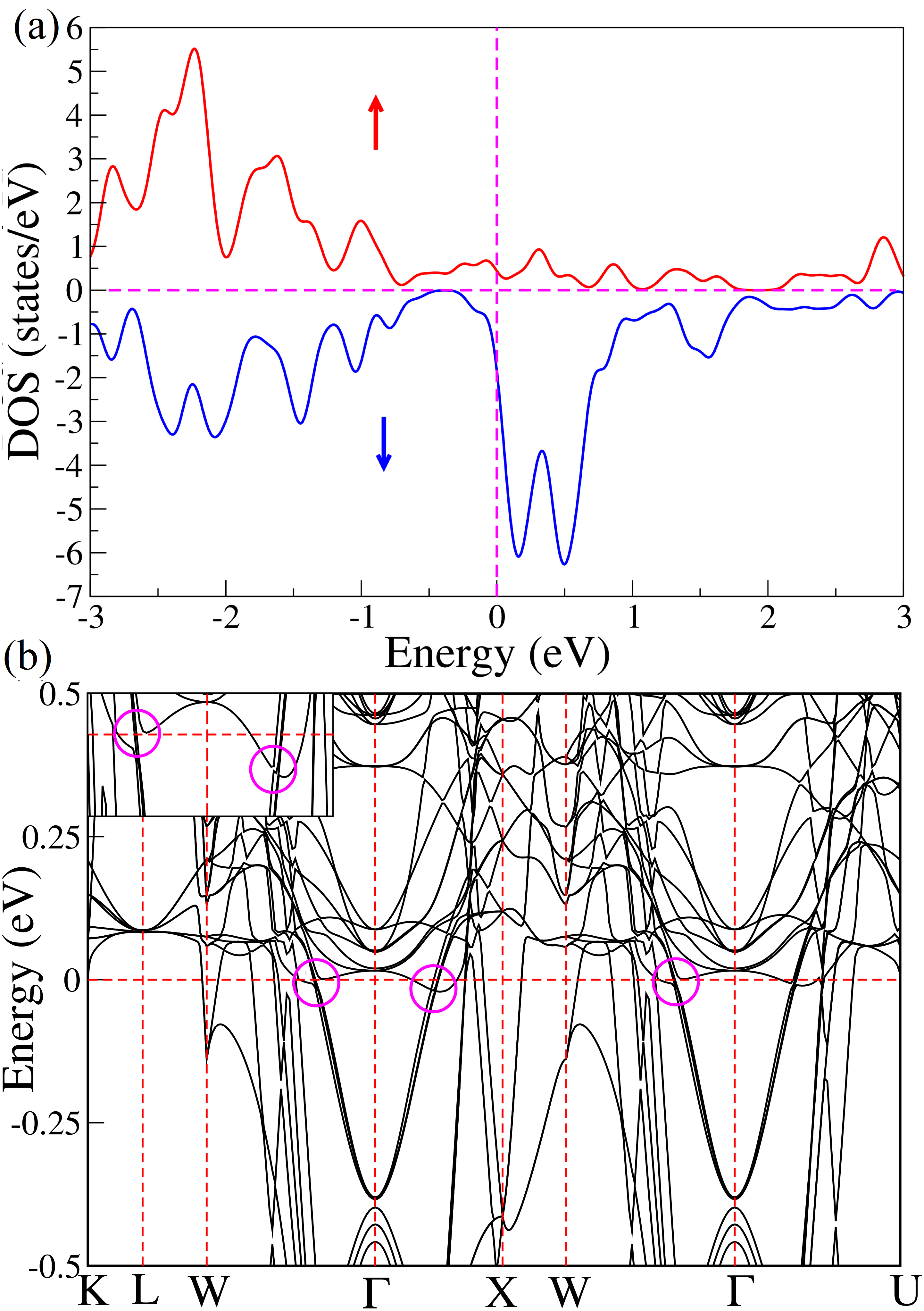}
\caption{\label{fig:Fig5.png}(a) The total density of state (DOS) of Co$_2$FeGe for on site Coulomb interaction $U$=0. Red and blue curves represent the total DOS for majority and minority spins. (b)The band structure of Co$_2$FeGe in the presence of SOC (gapped nodal lines are shown in circle and inset shows the enlarged view around gapped nodal line).}
\label{Fig5}
\end{figure}

%%%%%%%%%%%%%%%%%%%%%%%%%%%%%%%%%%%%%%%%%%
%\begin{figure}[t]
%\centering
%\includegraphics[width=0.5\textwidth]{Fig5.png}
%\caption{(a) spin-polarized band structure (b)The band structure of Co$_2$FeGe %(gapped nodal points are shown in circle and inset show the enlarged view of %gapped nodal points).}
%\label{Fig5}
%\end{figure}
%%%%%%%%%%%%%%%%%%%%%%%%%%%%%%%%%%%%
%%%%%%%%%%%%%%%%%%%%%%%%%%%%%%%%%%

%%%%%%%%%%%%%%%%%%%%%%%%%%%%%%%%%

In order to investigate the origin of AHE, we analysed the \si{\rho}\textsubscript{AH} versus \si{\rho}\textsubscript{xx} on a double logarithmic scale. A linear fitting was employed to determine the exponent $\beta$ according to the scaling relation \si{\rho}\textsubscript{AH}$\propto$ \si{\rho}\textsubscript{xx}$^\beta$ \cite{roy2020anomalous,wang2018large} shown in  Fig.\ref{fig3}(d). The exponent $\beta $ decides the dependency of \si{\rho}\textsubscript{AH} on \si{\rho}\textsubscript{xx}. According to well established theory of AHE,   $\beta $ =1  is for the AHE originates from the skew scattering mechanism and  $\beta $= 2  is for the AHE govern by scattering independent mechanism  \cite{nagaosa2010anomalous}.
 By this procedure, we found the exponent $\beta=1.75$, which indicates the Berry phase mechanism as dominant contribution in AHE. To find the value of intrinsic AHC, we have plotted \si{\rho}\textsubscript{AH} versus \si{\rho}\textsubscript{xx} (Inset of Fig.\ref{fig3}d) and fitted with equation
\begin{equation}
 \si{\rho}\textsubscript{AH} = \mbox{a}\si{\rho}\textsubscript{xx}+b\si{\rho}\textsubscript{xx}^2  
\end{equation}
Here  a and b are the skew scattering coefficient and intrinsic AHC, respectively. By this, we found a= 0.0004 and intrinsic AHC b $\sim$ 78 S/cm. 
The AHC due to extrinsic side jump contribution usually order of e$^2$/hc($\epsilon\textsubscript{SOC}$/E\textsubscript{F}), where $\epsilon\textsubscript{SOC}$ and E\textsubscript{F} are the spin-orbit interaction energy and Fermi energy respectively \cite{campbell1982transport}. For metallic ferromagnets  $\epsilon\textsubscript{SOC}$/E\textsubscript{F} is generally less than 10$^{-2}$ \cite{roy2020anomalous,liu2021anomalous} and hence the side jump contribution in AHC is too small or negligible in comparison to intrinsic AHC.
%This  gives the intrinsic AHC $\sim$ 78 S/cm.  
Further, to understand the microscopic origin of AHE, we need to look towards the variation of AHC with temperature and/or longitudinal resistivity. For this, we calculated Hall conductivity using tensor conversion formula
\begin{equation}
\si{\sigma}\textsubscript{H}=\frac{\si{\rho}\textsubscript{H}}{(\si{\rho}\textsubscript{xx}^2+\si{\rho}\textsubscript{H}^2)}
\end{equation}
Fig.\ref{fig4}a displays the field dependent Hall conductivity curves at indicated temperatures. The AHC  was calculated by zero field extrapolation of high field Hall conductivity data on y axis and found close to 100 S/cm at room temperature.  Fig.\ref{fig4}b shows  variation of  longitudinal conductivity and AHC with temperature.  AHC is nearly insensitive to temperature from several K to 300K, while the longitudinal resistivity shows explicit temperature dependence, fairly indicates the origin of AHE governed by the intrinsic mechanism \cite{liu2019magnetic,wang2018large,chen2019pressure}. Since the intrinsic AHE merely depends on the band structure of material so to get a better understanding of the origin of intrinsic AHE, we carried out first principles calculation on Co$_2$FeGe .

\begin{figure*}[t]
\centering
\includegraphics[width=0.9\textwidth]{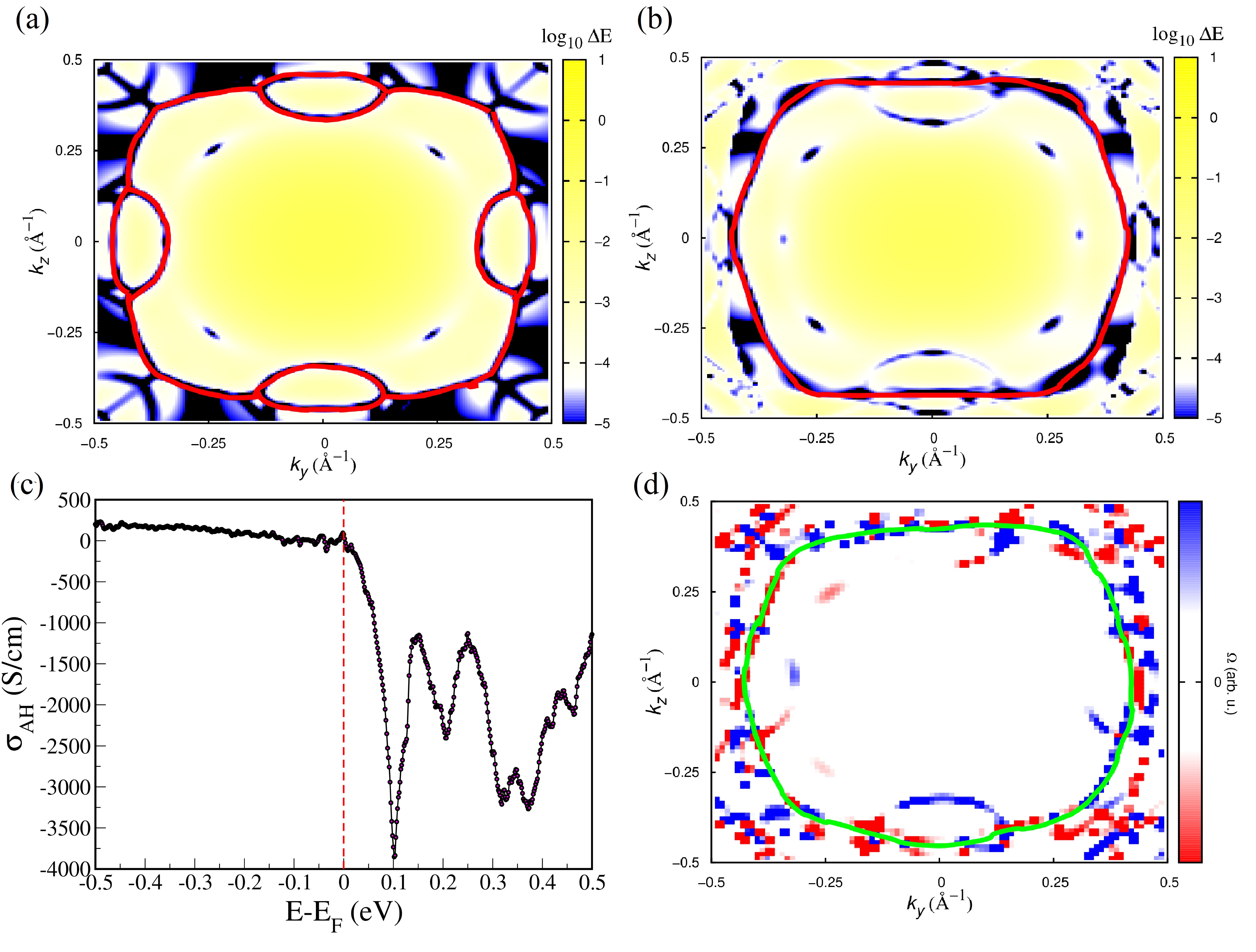}
%\centering
\caption{ Energy gap $\Delta{E}(\textit{k}_y,\textit{k}_z)$ is plotted in $\textit{k}_y$-$\textit{k}_z$ plane at $\textit{k}_x=0$ (a) without SOC (b)  with SOC.  Solid red lines represent gapless regions. (c) Energy ($E-E_F$) dependence of the AHC. (d) Berry curvature distribution in $\textit{k}_y$-$\textit{k}_z$ plane at $\textit{k}_x=0$.Solid red lines represent gapless regions.}
\label{fig6}
\end{figure*}

 Our ab-initio calculation for a magnetic moment suggests that Fe and Co have a large magnetic moment with $\mu_{Fe}$=2.84 $\mu_B$/f.u and $\mu_{Co}$=1.34 $\mu_B$/f.u respectively, whereas Ge has a small vanishing  magnetic moment.  The total magnetic moment per formula unit is 5.53 $\mu_B$, aligned along the (001) direction and the d-orbital of the transition atoms Fe and Co are the major contributors.
 The magnetic moment of full Heusler alloys generally follow the Slater Pauling (SP) rule\cite{faleev2017unified}; $M=Z-24$, where M and Z are magnetic moments and number of valance electrons, if the E$_F$ lies in the band gap of the minority spin states. For Co$_2$FeGe, as per SP rule the total magnetic moment should be 6 $\mu_B$/f.u. In some cases these systems may have gapless minority bands or  minority bands crossing the E$_F$, then a small deviation in total magnetic moment from the SP rule may be expected \cite{faleev2017unified}. The calculated magnetic moment for Co$_2$FeGe is in well agreement with the experimental value (5.38 $\mu_B$/f.u), but this value is smaller than the predicted by the SP rule. To understand the deviation of magnetic moment total density of state(DOS) is calculated and shown in Fig.\ref{Fig5}a. Total DOS of minority spin electrons have finite value at $E_F$, which indicates presence of gapless states at the $E_F$ or bands crossing $E_F$. Similar finite value of total DOS at $E_F$ is also reported in literature for onsite Coulomb interaction $U=0$ \cite{rai2012comparative,kumar2009first,hyun2018half}.
We also performed calculations considering onsite Coulomb repulsion $U$; 1.92 and 1.8 for the Co and Fe, respectively \cite{nawa2019exploring} and the calculated value of magnetic moment is $\approx 6.00$ $\mu_B$/f.u, which overestimates relative to experimental value, however, the magnetic moment is in agreement with literature \cite{rai2012comparative,balke2008rational,hyun2018half}. Therefore, we believe that $U= 0$ is more suitable parameter to explain the experimental findings. Similar conclusion is also reported for other Co$_2$-based Heusler alloys \cite{chang2016room,chang2017topological}.

This material is expected to show nodal line due to three relevant mirror planes \textit{m}$_x$(\textit{k}$_x$=0),\textit{m}$_y$(\textit{k}$_y$=0), \textit{m}$_z$(\textit{k}$_z$=0) in absence of SOC\cite{noky2020giant}. In presence of SOC the nodal lines will gap out according to the magnetization direction and introduce Berry curvature in the system \cite{noky2020giant,manna2018heusler}. To understand the topological aspects, the nodal lines in the electronic band structure are analysed in absence and presence of SOC. The band structure with SOC coupling is calculated from the DFT calculation shown in Fig.\ref{Fig5}b. The band crossings are gaped out just below or at the E$_F$ due to perturbation of SOC and these tiny gaps are shown inside the circle.  

We analyzed the spectrum of  tight binding model Hamiltonian calculated from the Wannier90~\cite{Pizzi_2020,marzari_97} calculation with the help of Wanniertool~\cite{WU2018405}. In Fig.\ref{fig6}a and Fig.\ref{fig6}b, we have shown energy gap between the lowest conduction band and the topmost valence band at \textit{k}$_x$=0 plane and gaps are shown on a logarithmic scale of color plot in absence as well as in the presence of SOC, respectively. The gaps smaller than 10$^{-4}$ eV are considered to be gapless. We noticed that  four semi-circular nodal lines appear in \textit{$k_y-k_z$} plane of first Brillouin zone (Fig.\ref{fig6}a) and in the presence of the SOC, the gapless semi-circular nodal lines are gaped out (Fig.\ref{fig6}b). This material shows an intrinsic AHC, which can be expressed within the framework of linear response theory of the Kubo formalism\cite{Gradhand_2012};

\begin{equation}
 \sigma_{\alpha\beta} = -{\frac{e^2}{\hbar} \sum_{n}\int\frac{d^{3}\textit{k}}{(2\pi)^3}\Omega^n_{\alpha\beta}(\textit{k})f_n(\textit{k})}  \end{equation}
 where Berry curvature $\Omega$ can be written as a sum over eigenstates\cite{xiao2010berry}.
\begin{eqnarray}
\Omega^n_{\alpha \beta} = i \sum_{n \neq n'} \frac{{\langle n|\frac{\partial H}{\partial R^\alpha}|n'\rangle} {\langle n'|\frac{\partial H}{\partial R^\beta}|n \rangle}-(\alpha\xleftrightarrow{}\beta)}{(\epsilon_n - \epsilon_{n^{'}})^2}
\end{eqnarray}
Here $\ket{n}$,$\epsilon_n$ and $\epsilon_{n'}$ are the energy eigenstate and eigenvalue of $n$ and $n'$ bands respectively. $H$, $f_n$ and $\Omega^n_{\alpha \beta}$ are the Hamiltonian, Fermi distribution function and Berry curvature respectively. The $\Omega^n_{\alpha \beta}$ is related to the change of electronic wave-function within the Brillouin zone. The detail of Berry curvature is in the appendix\ref{Appendix}.
 
In Fig.\ref{fig6}d, local Berry curvature is shown in $k_x=0$ plane of Brillouin zone.Intrinsic AHC is calculated using maximally localized Wannier orbitals using 101 × 101 × 101 k grid. The intrinsic AHC is proportional to the Brillouin zone summation of the Berry curvature over all occupied states and can be calculated using Eq.(3). We notice that the major contribution of AHC comes from neighborhood of nodal lines. The AHC as a function of $E-E_F$ is shown in Fig.\ref{fig6}c. The calculated AHC value at $E_F$ is found 77.29 S/cm, which well matches with the experimentally found intrinsic AHC 78.6 S/cm. We have also calculated the energy dependent AHC as shown in Fig.\ref{fig6}c and it is evident that Fermi level shift in Co$_2$FeGe will result into an increased  AHC.
 %However, a small difference between the theoretically and experimentally found  AHC due to small contribution of skew scattering.
%\section{Conclusion}
%\vspace*{3mm}

In conclusion, we have experimentally investigated the AHE in Co$_2$FeGe  Heusler alloy and  performed first principle calculations to understand the origin of intrinsic AHE. Experimentally AHC was found close to 100 S/cm at 300 K with an intrinsic contribution of 78.6 S/cm. Berry curvature calculations give AHC about 77.29 S/cm due to magnetization induced gapped nodal line near the E$_F$, which is in well agreement with the experimentally calculated intrinsic AHC.
%\section{Acknowledgements}
%\vspace*{3mm}

SS thanks Science and Engineering Research Board of India for financial support through the award of Ramanujan Fellowship (grant no: SB/S2/RJN-015/2017) and Early Career Research Award (grant no: ECR/2017/003186) for financial support. GKS thanks DST-INSPIRE scheme for fellowship. MK thanks DST for funding through grant no. CRG/2020/000754.
\par
\medskip
%\section*{Appendix}
%\vspace{-1mm}
%\label{appendix}
\appendix
\label{Appendix}
\renewcommand{\theequation}{A.\arabic{equation}}
\setcounter{equation}{0}
\section*{Appendix}
In the semi-classical limit, electrons moving in bands of the magnetic system require additional terms for the anomalous contribution to the conductivity and correction is done considering the wave packet rather than just particle movement \cite{Gradhand_2012}. Wave packet can be written as
 \begin{equation}
 W_{\textit{k}_c,\mathbf{r}_c}(\mathbf{r},t)=\sum_{\textit{k}} a_\textit{k}(\textit{k}_c,t)\psi_{n\textit{k}}(\mathbf{r})
 \end{equation}
 where, $\psi_{n\textit{k}}(\mathbf{r}=e^{i\textit{k}\cdot\mathbf{r}}u_{n}(\mathbf{r},\textit{k})$. A wave packet is strongly centered at $\textit{k}_c$ in the Brillouin zone. $\mathbf{r}_c$ is the spatial center position of a wave packet. Here, $\mathbf{r_c}$ is defined by-
\begin{equation}
\mathbf{r}_c=\bra{W_{\textit{k}_c,\mathbf{r}_c}}\mathbf{r}\ket{W_{\textit{k}_c,\mathbf{r}_c}}
 \end{equation}
and $a_\textit{k}(\textit{k}_c,t)$,the  phase of the weighting
function can be given as
\begin{equation}
a_\textit{k}(\textit{k}_c,t)=|a_\textit{k}(\textit{k}_c,t)|e^{i(\textit{k}-\textit{k}_c)\cdot\mathbf{A}_n(\textit{k}_c)-i\textit{k}\cdot\mathbf{r}_c}
\end{equation}

\begin{eqnarray}
\mathbf{A}_n(\textit{k}_c)=i\int_{u.c.}\mathbf{d}^{3}r u^{*}_{n}(\mathbf{r},\textit{k}_c)\nabla_{\textit{k}}u_{n}(\mathbf{r},\textit{k}_c)
\end{eqnarray}
$\mathbf{A}_n(\textit{k}_c)$ is the Berry connection.Using the wave packet of equation, one can write the Lagrangian and thereafter anomalous velocity can be calculated solving the Lagrangian. Anomalous velocity can be given as\cite{Gradhand_2012}

\begin{eqnarray}
v_{n}=\frac{e}{\hbar}E\times\Omega_n(k_c)
\end{eqnarray}

 where $E$ is an electric field and $\Omega{_n}(k_c)$ is Berry curvature. The Berry curvature of a solid-state material in tight binding limit can be written as  $\Omega_n(k)=\Omega_n^k(k)+\Omega_n^r(k)$. For non-degenerate energy bands, these are defined as

\begin{eqnarray}
&& \Omega_{n}^\textit{k}(\textit{k})= -i\sum_{m\neq n} \bigg{[}\frac{C_{n}^{\dagger}(\textit{k})\nabla_{\textit{k}}H(\textit{k})C_{m}(\textit{k})}{(\epsilon_{n\textit{k}} - \epsilon_{m\textit{k}})^{2}}\nonumber \\
&&  \times C_{m}^{\dagger}(\textit{k})\nabla_{k}H(\textit{k})C_{n}(\textit{k}) \bigg{]}
\end{eqnarray}

\begin{eqnarray}
&&  \Omega_{n}^{r}(\textit{k})=\sum_{m\neq n} 2Re \bigg{[} \frac{C_{n}^{\dagger}(\textit{k})\nabla_{\textit{k}}H(\textit{k})C_{m}(\textit{k})}{(\epsilon_{n\textit{k}} - \epsilon_{m\textit{k}})} \nonumber \\
&&  \times C_{m}^{\dagger}(\textit{k})\mathbf{r} C_{n}(\textit{k}) \bigg{]}
\end{eqnarray}
 Where $\epsilon_{nk}$ and $C^n_{\textit{k}}$ are energy of $n^{th}$ band and coefficient of wave function respectively. $H(\textit{k})$ is a Hamiltonian. However, for the degenerate band, the $\Omega_{n}(k_c)$ can be calculated as in equations 73 and 74 of ref \cite{Gradhand_2012}.
\bibliography{references}
\end{document}